\documentstyle[12pt]{article}
\title{\bf Classical Euclidean wormhole solutions in Palatini $f(\tilde{R})$ cosmology}
\author{F. Darabi \thanks{f.darabi@azaruniv.edu} \\
{\small Department of Physics, Azarbaijan Shahid Madani University, 53714-161 Tabriz, Iran .}}
\begin{document}
\def\pp{{\, \mid \hskip -1.5mm =}}
\def\cL{{\cal L}}
\def\beq{\begin{equation}}
\def\eneq{\end{equation}}
\def\bea{\begin{eqnarray}}
\def\enea{\end{eqnarray}}
\def\tr{{\rm tr}\, }
\def\nn{\nonumber \\}
\def\e{{\rm e}}
\maketitle
\begin{abstract}
We study the classical Euclidean wormholes in the context of extended theories of gravity. With no loss of generality, we use the dynamical equivalence between $f(\tilde{R})$ gravity and scalar-tensor theories to construct a point-like Lagrangian in the flat FRW space time. We first show the dynamical equivalence between Palatini $f(\tilde{R})$ gravity and the Brans-Dicke theory with self-interacting potential, and then show the dynamical equivalence between the Brans-Dicke theory with self-interacting potential and the minimally coupled O'Hanlon theory. We show the existence of new Euclidean wormhole solutions for this O'Hanlon theory and, for an special case, find out the corresponding form of $f(\tilde{R})$
having wormhole solution. For small values of the Ricci scalar, this $f(\tilde{R})$ is in agreement with the wormhole solution obtained for higher order
gravity theory $\tilde{R}+\epsilon \tilde{R}^2\:,\: \epsilon<0$.
\end{abstract}

\newpage

\section{Introduction}

There are two kinds of classical wormholes: Lorentzian and Euclidean. 
Lorentzian wormholes are known as the vacuum solutions to the Lorentzian
Einstein field equations, such as Schwarzschild wormholes or Einstein-Rosen bridges. These are {\it real bridges} between different areas of spacetime. According to Matt Visser's definition of Lorentzian wormholes: If a Minkowski spacetime contains a compact region $\Omega$, and if the topology of $\Omega$ is of the form $\Omega \sim R \times \Sigma$, where $\Sigma$ is a three-manifold of the nontrivial topology, whose boundary has topology of the form $d\Sigma \sim S^2$, and if, furthermore, the hypersurfaces $\Sigma$ are all spacelike, then the region $\Omega$ contains a quasipermanent intra-universe wormhole \cite{Matt}.

On the other hand, Euclidean wormholes have been studied mainly as
instantons, namely solutions of the classical Euclidean Einstein field
equations. Euclidean wormholes are usually considered as Euclidean metrics
that consist of two asymptotically flat regions connected by a
narrow throat (handle). In general, such wormholes can represent {\it quantum
tunneling} between different areas of spacetime having generally different
topologies. They are possibly useful in understanding black hole evaporation \cite{Haw}; in allowing nonlocal connections that could determine fundamental constants; and in vanishing the cosmological constant $\Lambda$ \cite{Col1}-\cite{Col3}.
They are even considered as an alternative to the Higgs mechanism
\cite{Mig}. Consequently, such solutions are of particular importance in
the study of quantum aspects of gravity.
The reason why classical wormholes may exist is related to the
implication of a theorem of Cheeger and Glommol \cite{Chee} which
states that a {\it necessary} (but not sufficient) condition for a classical wormhole to exist is that the eigenvalues of the Ricci tensor be negative {\it somewhere} on the manifold \cite{Gid1}. 

The Lorentzian wormholes have been extensively studied in the modified theories of gravity, like Brans-Dicke gravity \cite{Lobo0}, \cite{Lobo1}, \cite{Lobo2} modified teleparralel gravity \cite{Lobo3}, and $f(R)$ gravity \cite{Lobo4}.
However, to the author's knowledge, the study of Euclidean wormholes, in modified gravities, has not been received much attention. Motivated by this
ignorance, we are interested in the study of Euclidean wormholes in $f(R)$ gravity.
There are two formalisms to study $f(R)$ gravity: metric $f(R)$ gravity
and Palatini $f(R)$ gravity \cite{Odin1}-\cite{Capoz}. The first is the standard metric formalism in which the field equations are derived by the variation of the action with respect to the metric tensor. The second is the Palatini formalism in which the metric and connection are treated as independent variables in the variation of action. These two formalisms generally give rise to different field equations for a non-linear action,
however, it is well-known that the solutions of Palatini $f(R)$ gravity represent
sub-class of solutions of metric $f(R)$ gravity \cite{alle}-\cite{alle2}.
In fact, every Palatini model of gravity possess its purely metric counterpart which leads to fourth order field equations. In a recent paper \cite{Darabi}, we have considered four types of actions: metric-Jordan, Palatini-Jordan,  metric-Einstein and Palatini-Einstein. The symmetry between Jordan and Einstein frames is a conformal symmetry which corresponds to the appearance or vanishing of kinetic terms. On the other hand, transition from metric-Jordan action to Palatini-Jordan action requires the appearance of kinetic term, while the transition from metric-Einstein action to the Palatini-Einstein action requires the vanishing of kinetic term. In both transitions from ``metric-Jordan'' to ``metric-Einstein and Palatini-Jordan'' actions we have appearance of kinetic terms and in both transitions from ``metric-Einstein and Palatini-Jordan'' actions to ``Palatini-Einstein'' actions we have vanishing of kinetic term. Jordan and Einstein frames are dynamically equivalent from the conformal symmetry viewpoint. Although the  metric and Palatini formalisms are also connected through a conformal transformation, they don't seem to be dynamically equivalent. Metric-Jordan action differs from Palatini-Jordan action with a dynamical {\it advanced} kinetic term. In the same manner, metric-Einstein action differs from Palatini-Einstein action with a dynamical {\it retarded} kinetic term. However, the Palatini-Jordan action when reduced to Palatini-Einstein action takes the same form as the metric-Jordan action, namely it becomes of the O'Hanlon type action where the dynamics is completely endowed by the self interacting potential. On the other hand, metric-Einstein action and Palatini-Jordan action represent the same dynamical features in that both have a dynamical kinetic term plus a potential. In conclusion, for each map between Jordan and Einstein frames there exists a corresponding map between Palatini and metric formalisms. In the same way, for each map connecting two O'Hanlon type actions, namely metric-Jordan and Palatini-Einstein action, there exists a map which connects Palatini-Jordan action with metric-Einstein action. Therefore, the apparent differences between Palatini and metric formalisms strictly depend on the representation while the number of degrees of freedom is preserved. This means that the dynamical content of both formalism is identical. In fact, the metric and Palatini formalisms become non-equivalent on-shell in the presence of matter \cite{alle2}.  However, an advantage of Palatini formulation rely on second order field equations which turns out to be more easy to solve. In this sense the Palatini formalism is more easy to handle and simpler to analyse than the corresponding metric formalism \cite{alle2}. Moreover, if one considers the Einstein frame as the physical
frame, then the Palatini formalism is more convenient than the metric formalism
in the present study of Euclidean wormholes. This is because unlike the metric
formalism whose action is turned into the scalar tensor theory in the Jordan
frame, the action in the Palatini formalism may cast in the form of Einstein frame. Motivated by the above discussion, we aim to study the Euclidean
wormholes in Palatini $f(R)$ gravity.

The work in the present paper is similar in spirit to the works done by the
authors in Refs.\cite{Fuk1}-\cite{Fuk6}: We use the dynamical equivalence between Palatini $f(\tilde{R})$ gravity \cite{Buch} and scalar-tensor theories to construct a point-like Lagrangian in the flat Friedmann-Robertson-Walker (FRW) universe \cite{Sot,Ros}. In so doing, the well known dynamical equivalence between Palatini $f(\tilde{R})$ gravity and the Brans-Dicke theory with self-interacting potential is briefly reviewed \cite{Cap,Cap1,Darabi1}. Then, the well known dynamical equivalence between the Brans-Dicke theory with self-interacting potential and the minimally coupled O'Hanlon theory is studied, where the dynamics is completely endowed by the self interacting potential. Finally, the existence of new Euclidean wormhole solutions for this O'Hanlon theory is shown and the possible corresponding forms of $f(\tilde{R})$ in Palatini formalism are obtained.

\section{Dynamical equivalence between Palatini $f(\tilde{R})$ gravity and minimally coupled O'Hanlon theory}

The action of the Palatini $f(\tilde{R})$ theories takes the following form
\begin{eqnarray}\label{eq.1}
S=\frac{1}{2k}\int
d^{4}x\sqrt{-g}f(\tilde{R}),
\end{eqnarray}
where $f(\tilde{R})$ is a function of $\tilde{R}=g^{\mu\nu}
R_{\mu\nu}(\tilde{\Gamma})$ and
$\tilde{\Gamma}^{\lambda}_{\mu\nu}$ is the connection. This action depends on the two dynamical variables, namely metric and connection. Variation of Eq.(\ref{eq.1}) with respect to the metric leads to
\begin{eqnarray}\label{eq.2}
f'(\tilde{R})\tilde{R}-\frac{1}{2} f(\tilde{R})g_{\mu\nu}= 0,
\end{eqnarray}
where $f'(\tilde{R})=df/d\tilde{R}$. The trace of Eq.(\ref{eq.2}) is
\begin{eqnarray}\label{eq.3}
f'(\tilde{R})\tilde{R}-2f(\tilde{R})=0,
\end{eqnarray}
and the variation of Eq.(\ref{eq.1}) with respect to the connection gives
\begin{eqnarray}\label{eq.4}
(\sqrt{-g}f'(\tilde{R})g^{\mu\nu})_{;\lambda}=0,
\end{eqnarray}
where $;$ denotes covariant derivative. Therefore, the connection is compatible with the new metric
$h_{\mu\nu}=f'(\tilde{R})g_{\mu\nu}$ and we
obtain
\begin{eqnarray}\label{eq.5}
\tilde{R}=R+\frac{3}{2f'(\tilde{R})}
\partial_{\lambda}f'(\tilde{R})\partial^{\lambda}f'(\tilde{R})-\frac{3}{f'(\tilde{R})}\Box f'(\tilde{R}),
\end{eqnarray}
where $R$ is Ricci scalar constructed from the Levi-Civita connection
of the metric $g_{\mu\nu}$. One can easily
verify that the action (\ref{eq.1}) is dynamically equivalent to \cite{Sot}
\footnote{Using a general theory with a divergence-free current, one can also demonstrate the equivalence between the action (\ref{eq.6})
in Palatini formalism and the following action
$$
S=\frac{1}{2k}\int d^{4}x \sqrt{-g}(\Phi R-V(\Phi)),
$$
in metric formalism of $f(R)$ gravity \cite{Darabi1}. In fact, considering a simple divergence theory, and a suitably defined current in terms of the field $\Phi$, it is possible
to generalize the conformal equivalence between metric and Palatini formalisms
by a conservation equation of this current.}
\begin{eqnarray}\label{eq.6}
S=\frac{1}{2k}\int d^{4}x \sqrt{-g}(\Phi R+\frac{3}{2\Phi}
\Phi_{;\mu} \Phi^{;\mu} -V(\Phi)), 
\end{eqnarray}
where $\Phi=f'(\tilde{R})$ ,
$V(\Phi)=\chi(\Phi)\Phi-f(\chi(\Phi))$ and $\tilde{R}=\chi(\Phi)$. 
This is the well-known action of  Brans-Dicke theory with the
coupling parameter equal to $-\frac{3}{2}$. Using the redefinition
$\Phi\equiv\varphi^{2}$ the action (\ref{eq.6}) takes the following
form
\begin{eqnarray}\label{eq.7}
S=\frac{1}{2k}\int d^{4}x \sqrt{-g}(\varphi^2 R+6
\varphi_{;\mu} \varphi^{;\mu} -V(\varphi)).
\end{eqnarray}
This action is dynamically equivalent to 
\begin{eqnarray}\label{eq.8}
S=\frac{1}{2k}\int d^{4}x \sqrt{-g}(F(\varphi) R+\frac{1}{2}
\varphi_{;\mu} \varphi^{;\mu} -U(\varphi)),
\end{eqnarray}
where $F(\varphi)=\frac{1}{12}\varphi^2$ and $ U(\varphi)=\frac{1}{12}V(\varphi)$.
A conformal transformation of the following type
\cite{Cap}, \cite{Cap1}
\begin{eqnarray}\label{eq.9}
\bar{g}_{\mu \nu}=e^{2\sigma}g_{\mu \nu},
\end{eqnarray}
results in the Lagrangian density in the Einstein frame
\begin{equation}\label{eq.12}
\sqrt{-g}(F R+\frac{1}{2}\varphi_{;\mu} \varphi^{;\mu} -U)=
\sqrt{-\bar{g}}\left(\frac{1}{2}\bar{R}
+3\Box_{\bar{\Gamma}}\sigma+\frac{3F_{\varphi}^2-F}{4F^2}\varphi_{;\alpha}\varphi^{\alpha}_{;}
-\frac{U}{4F^2}\right).
\end{equation}
By introducing a new scalar field $\bar{\varphi}$ and the potential $\bar{U}$,
respectively, defined by
\begin{equation}\label{eq.13}
\bar{\varphi}_{;\alpha}=\sqrt{\frac{3F_{\varphi}^2-F}{4F^2}}\varphi_{;\alpha},
\:\:\:\:\: \bar{U}(\bar{\varphi}(\varphi))=\frac{U}{4F^2},
\end{equation}
we obtain \cite{Cap}
\begin{equation}\label{eq.14}
\sqrt{-g}(FR+\frac{1}{2}g^{\mu \nu}\varphi_{;\mu} \varphi^{;\mu}-U)=
\sqrt{-\bar{g}}\left(\frac{1}{2}\bar{R}
+\frac{1}{2}\bar{\varphi}_{;\alpha}\bar{\varphi}^{\alpha}_{;}
-\bar{U}\right).
\end{equation}
If we put $F(\varphi)=\frac{1}{12}\varphi^2$ in the first definition
in Eq.(\ref{eq.13}) we obtain $\bar{\varphi}_{;\alpha}=0$ which leads to the following action
\begin{equation}\label{eq.15}
{\cal S}=\int d^4x \sqrt{-\bar{g}}\left(\frac{1}{2}\bar{R}
-\bar{U}\right),
\end{equation}
which is known as the O'Hanlon action in the Einstein frame where the dynamics is completely endowed by the self interacting potential \cite{Ohan}.

\section{Classical Euclidean wormholes in O'Hanlon theory}

In this section, we
look for the wormhole solutions in the system (\ref{eq.15}) of minimally coupled scalar field with the lagrangian density
\begin{equation}\label{eq.16}
{\cal L}=\frac{1}{2}\bar{R}
-\bar{U}.
\end{equation}
where $U(\bar{\varphi})$ is a self-interacting potential. We do not specify {\it a priori} the form of the potential, and by analyzing the field equations and the corresponding wormhole solutions we may use the conformal equivalence discussed in the previous section to go in the opposite direction and obtain the corresponding wormhole solutions in Palatini $f(R)$ gravity.
\\
The Einstein equations of motion are obtained as
\begin{equation}\label{eq.17}
\bar{R}_{\mu \nu}=\bar{T}_{\mu \nu}-\frac{1}{2}\bar{g}_{\mu \nu}\bar{T},
\end{equation}
where the energy-momentum tensor and its trace are given respectively by
\begin{equation}\label{eq.18}
\bar{T}_{\mu \nu}=\bar{g}_{\mu
\nu}\bar{U}(\bar{\varphi}),
\end{equation}
\begin{equation}\label{eq.19}
\bar{T}=4\bar{U}(\bar{\varphi}).
\end{equation}
Putting (\ref{eq.18}) and (\ref{eq.19}) in Eq.(\ref{eq.17}) we obtain 
\begin{equation}\label{eq.20}
\bar{R}_{\mu \nu}=-\bar{g}_{\mu
\nu}\bar{U}(\bar{\varphi}).
\end{equation}
It is seen that for the positive definite Euclidean metric $\bar{g}_{\mu \nu}$ the Ricci tensor $\bar{R}_{\mu \nu}$ has negative eigenvalues if and only if the potential $\bar{U}(\bar{\varphi})$ is positive, and consequently wormhole solutions may exist in this system
if and only if the following two conditions hold
\begin{equation}\label{eq.21}
\bar{U}(\bar{\varphi})>0,
\end{equation}
\begin{equation}\label{eq.22}
-\bar{g}_{\mu
\nu}\bar{U}(\bar{\varphi})<0.
\end{equation}
For the flat Friedmann-Robertson-Walker universe the Euclidean
metric $\bar{g}_{\mu \nu}$ is written as
\begin{equation}\label{eq.23}
dS^2=dt^2+a^2(t)d^2\Omega_3,
\end{equation}
where $d^2\Omega_3$ is the line element on the three-sphere. 
The Euclidean field equation for the variable $a$ is obtained as
\begin{equation}\label{eq.24}
\dot{a}^2=1-a^2\bar{U}(\bar{\varphi}),
\end{equation}
where an overdot denotes $d/dt$.
Now, we look for the wormhole solutions for the equation (\ref{eq.24}). 
It is generally believed that a wormhole has two asymptotically flat regions
connected by a throat at which $\dot{a}=0$ and it is described by an expression of the form 
\begin{equation}\label{eq.25}
\dot{a}^2=1-\frac{C}{a^n},
\end{equation}
where $C$ is a positive constant. In order to have an asymptotic Euclidean wormhole it is necessary that $\dot{a}^2$ remains positive at large $a$, and this requires $n>0$. Comparison of the equations (\ref{eq.24}) and (\ref{eq.25}) shows that it is possible to choose a suitable form of the
potential $\bar{U}(\bar{\varphi})$ so that Eq.(\ref{eq.24}) represents a
wormhole. Therefore, the existence of wormholes for O'Hanlon theory is established. Now, we wish to look for the corresponding wormholes in the Palatini $f(\tilde{R})$
theory. In so doing, we may rewrite the potential in the following form
\begin{equation}\label{eq.26}
\bar{U}(\bar{\varphi})=\frac{U}{4F^2}=3\frac{V(\varphi)}{\varphi^4}
=3\frac{\tilde{R}\Phi-f(\tilde{R})}{\Phi^2}
=3\frac{\tilde{R}f'-f(\tilde{R})}{f'^2}.
\end{equation}
In order this potential, after inclusion in Eq.(\ref{eq.24}), could represent a wormhole we should take the following equation
\begin{equation}\label{eq.27}
\frac{\tilde{R}f'-f(\tilde{R})}{f'^2}=\frac{C}{a^{n+2}}.
\end{equation}
One may wish to rewrite this equation as a first order differential equation for $f$ as a function of $a$ or $\tilde{R}$. In both cases we need to express
the Ricci scalar $\tilde{R}$ in terms of the scale factor $a$. 
Using the metric (\ref{eq.23}), the scalar
curvature is obtained as
\begin{equation}\label{eq.28}
\tilde{R}=6\left[\frac{\ddot{a}}{a}+\frac{\dot{a}^{2}}{a^{2}}\right].
\end{equation}
If we are to consider (\ref{eq.25}) as the wormhole solution, then the Ricci
scalar should take the following form
\begin{equation}\label{eq.29}
\tilde{R}=6\left[\frac{C}{a^{n+2}}(\frac{n}{2}-1)+\frac{1}{a^{2}}\right],
\end{equation}
where the values of $n$ and $C$  may fix the sign of the Ricci scalar.

It is of valuable practice to obtain the wormhole solutions in $f(\tilde{R})$
theory whose defining equations are the same as the well-known wormhole
solutions in conventional Einstein-Hilbert theory.  For the sake of simplicity we first take $n=2$. This case corresponds to
the typical known wormhole of conformal scalar field coupled with Einstein-Hilbert
action \cite{Hall1,Hall2,Hall3}.
The Ricci scalar and $f'$ then become respectively
\begin{equation}\label{eq.30}
\tilde{R}=\frac{6}{a^2},
\end{equation}
\begin{equation}\label{eq.31}
f'=\frac{df}{d\tilde{R}}=\frac{df}{da}\frac{da}{d\tilde{R}}=-\frac{a^3}{12}\frac{df}{da}.
\end{equation}
Putting Eqs.(\ref{eq.30}) and (\ref{eq.31}) in (\ref{eq.27}) we obtain the
following first order differential equation 
\begin{equation}\label{eq.32}
a^6\left(\frac{df}{da}\right)^2+Aa^5\left(\frac{df}{da}\right)+2Aa^4 f=0.
\end{equation}
We may also use (\ref{eq.32}) and its derivative $d/da$ to obtain the second order differential equation 
\begin{equation}\label{eq.33}
Aa^5\frac{d^2f}{da^2}+2a^6\frac{df}{da}\frac{d^2f}{da^2}+2a^5\left(\frac{df}{da}\right)^2+3Aa^4\frac{df}{da}=0, \end{equation}
where $A=\frac{72}{C}$. 
Alternatively, we may use Eq.(\ref{eq.30}) in Eq.(\ref{eq.27})
to express $a$ in terms of $\tilde{R}$. In doing so, we obtain the following
first order differential equation 
\begin{equation}\label{eq.34}
\tilde{R}^2\left(\frac{df}{d\tilde{R}}\right)^2-\tilde{R}\left(\frac{df}{d\tilde{R}}\right)+f=0. \end{equation}
Equivalently, using the derivative $d/d\tilde{R}$ of Eq.(\ref{eq.34}) we
obtain the following second order differential equation for $f$ in terms of $\tilde{R}$
\begin{equation}\label{eq.35}
\frac{d^2f}{d\tilde{R}^2}\left(1-2\tilde{R}\frac{df}{d\tilde{R}}\right)-2\left(\frac{df}{d\tilde{R}}\right)^2=0. \end{equation}
Since we are interested in the explicit function $f(\tilde{R})$, we consider
Eq.(\ref{eq.35}) whose solution may be obtained either in the parametric form 
\begin{equation}\label{eq.36}
\left\{ \begin{array}{ll} f(T)=\frac{1}{4}-\frac{1}{4}\ln(T)^2-\frac{C}{2}\ln(T)-\frac{1}{4}C^2, \\
\\
R(T)=\frac{1}{2T}[1-sgn[\ln(T)+C]\ln(T)-sgn[\ln(T)+C]C,
\\
\end{array}
\right.
\end{equation}
\begin{equation}\label{eq.37}
\left\{ \begin{array}{ll} f(T)=\frac{1}{4}-\frac{1}{4}\ln(T)^2-\frac{C}{2}\ln(T)-\frac{1}{4}C^2, \\
\\
R(T)=\frac{1}{2T}[1+sgn[\ln(T)+C]\ln(T)+sgn[\ln(T)+C]C,
\\
\end{array}
\right.
\end{equation}
with $C$ being a constant and ``$sgn$" denoting the {\it sign function}, or in the explicit form\footnote{The LambertW function
satisfies
$$
LambertW(x)\exp(LambertW(x))=x,
$$
and it has infinite number of branches for each (nonzero) value of $x$ while
exactly one of these branches is analytic at zero. For small values of $x$
we have $LambertW(x) \simeq x$.}
\begin{equation}\label{eq.38}
f( \tilde{R}) =-\frac{1}{4}\, \left[ {\it LambertW} \left( -2C_1\,\tilde{R}{\it } \right)  \right] ^{2}-\frac{1}{2}\,{\it LambertW} \left( -2\,C_1\tilde{R}{\it }
 \right) +{\it C_2},
\end{equation}
where $C_1$ and $C_2$ are constants and {\it LambertW} is the LambertW function.
For small values of the Ricci scalar $\tilde{R}$ we obtain, modula the constant
$C_1$, the following form 
\begin{equation}\label{eq.39}
f( \tilde{R}) \simeq \tilde{R}-C_1\tilde{R}^2+{\it \frac{C_2}{C_1}}.
\end{equation}
It is worth noting that the above solution, modula the constant term ${C_2}/{C_1}$, is the wormhole found by {\it Fukutaka et al} in the higher order
gravity theory $\tilde{R}+\epsilon\tilde{R}^2,\:\: \epsilon<0$, for closed Friedman-Robertson-Walker universe \cite{Fuku}. This interesting agreement
confirms the correctness of the general form $f( \tilde{R})$ given by Eq.(\ref{eq.38}) for which we expect classical Euclidean wormholes.

Next, we may take $n=4$ which corresponds to the {\it axion} field as the
matter source coupled with Einstein-Hilbert action that leads to the Giddings-Strominger
wormhole \cite{Gid1}. The Ricci scalar and $f'$ then become respectively
\begin{equation}\label{eq.40}
\tilde{R}=6\left[\frac{C}{a^6}+\frac{1}{a^2}\right],
\end{equation}
\begin{equation}\label{eq.41}
f'=\frac{df}{d\tilde{R}}=\frac{df}{da}\frac{da}{d\tilde{R}}=-\frac{1}{6}\frac{df}{da}\left[\frac{a^7}{6C+2a^4}\right].
\end{equation}
Putting Eqs.(\ref{eq.40}) and (\ref{eq.41}) in (\ref{eq.27}) we obtain the
following first order differential equation 
\begin{equation}\label{eq.42}
Aa^8\left(\frac{df}{da}\right)^2+a(C+a^4)(6C+2a^4)\frac{df}{da}+(6C+2a^4)^2f=0. \end{equation}
We may also use (\ref{eq.42}) and its derivative $d/da$ to obtain the second order differential equation 
\begin{eqnarray}\label{eq.43}
a(C+a^4)(6C+2a^4)\frac{d^2f}{da^2}+2Aa^8\frac{df}{da}\frac{d^2f}{da^2}&+&8Aa^7\left(\frac{df}{da}\right)^2
\\
\nonumber
&+&(42C^2+64Ca^4+22a^8)\frac{df}{da}+a^3(96C+32a^4)f=0.
\end{eqnarray}
where $A=\frac{36}{C}$. Unfortunately, neither the first order nor the second
order differential equations gives exact solutions (at least using the available mathematical software like Maple). 

Alternatively, if we wish to use Eq.(\ref{eq.40}) to express $a$ in terms
of $\tilde{R}$ and construct a differential equation like Eqs.(\ref{eq.34}), (\ref{eq.35}), then we obtain more complicate differential equations with no exact solution, so we ignore to follow seriously this case.

\section*{Conclusions}

Every Euclidean wormhole solution is of particular importance from macroscopic and microscopic points of view. In particular, very small Euclidean wormholes are studied as instantons, namely the saddle points in the Euclidean path integrals. So, one can use them to give a semi-classical treatment in the dilute wormhole approximation where the interaction between the large scale ends of wormholes is neglected. On the other hand, since the black holes evaporate in theories with reasonable matter contents, then new wormhole solutions may provide new contributions for black hole evaporation. In the same way, new wormhole solutions are supposed to play their own important roles in vanishing the cosmological constant. Taking into account the importance of new wormhole solutions and motivated by the existence of Euclidean wormhole solutions for some higher-order gravity theories we have studied the classical Euclidean wormhole solutions for modified general $f(\tilde{R})$ theories of gravity in Palatini formalism. We used a well known dynamical equivalence between $f(\tilde{R})$ gravity and minimally coupled O'Hanlon theory. We realized the existence of new Euclidean wormhole solutions for this O'Hanlon theory, and for an special case we obtained the corresponding (wormhole) form of $f(\tilde{R})$ which, for small $\tilde{R}$, is in agreement with the wormhole solution obtained for higher order gravity theory $\tilde{R}+\epsilon \tilde{R}^2\:,\: \epsilon<0$. 

In general, such wormholes in $f(\tilde{R})$ gravity can represent the same characteristic features as studied in GR. They may be used in a) the description of quantum amplitude for tunneling between different areas of spacetime, b) realizing black hole evaporation, c) allowing nonlocal connections in determination of fundamental constants, d) vanishing the cosmological constant, e) semi-classical treatment in the dilute wormhole approximation. Moreover, such wormholes may be used to understand weather they possibly affect the dynamical feature of $f(\tilde{R})$ gravity which effectively corresponds to the problem of dynamical dark energy.

As mentioned before, it is well-known fact that the metric and Palatini  theories of extended gravity become non-equivalent on-shell in the presence of matter, however, the solutions of Palatini $f(\tilde{R})$ gravity represent sub-class of solutions of metric $f(R)$ gravity in pure theory of $f(R)$
\cite{alle}-\cite{alle2}. Also, one can show the equivalence between metric
and Palatini formalisms in pure $f(R)$ gravity using divergence free currents
\cite{Darabi1}. As a result, since we have considered only purely gravitational case, the wormhole solutions of the present paper may be used to obtain the corresponding solutions of metric $f(R)$ gravity. In other words, the reported
results here may be framed in a more general context where Palatini and metric
approaches for extended theories of gravity are considered \cite{Ext}.

The solutions here are obtained for the isotropic FRW cosmology having one
scale factor representing the wormhole throat. The extension to other anisotropic Bianchi cosmological models with three different scale factors is also an interesting activity. To this end, one has to first establish the physical interpretation of wormhole solutions having more than one throat, and then
look for such possible wormhole like solutions in anisotropic Bianchi cosmological models.

\section*{Acknowledgment}
The author would like to thank the anonymous referee for enlightening comments on this paper. This research was supported by a grant/research fund Number 217/D/2659 from Azarbaijan University of Tarbiat Moallem, Tabriz, Iran.

\end{document}